\documentclass{article}

\usepackage{microtype}
\usepackage{graphicx}
\usepackage{subfigure}
\usepackage{booktabs} % for professional tables
\usepackage{multirow}

\usepackage{hyperref}

\usepackage[accepted]{icml2025}

\usepackage{amsmath}
\usepackage{amssymb}
\usepackage{mathtools}
\usepackage{amsthm}

\usepackage[capitalize,noabbrev]{cleveref}

\theoremstyle{plain}
\newtheorem{theorem}{Theorem}[section]

\theoremstyle{definition}

\theoremstyle{remark}

\usepackage[textsize=tiny]{todonotes}

\icmltitlerunning{MUFFIN: Multi-band Frequency Reconstruction for Neural Psychoacoustic Coding}

\begin{document}

\twocolumn[
\icmltitle{Multi-band Frequency Reconstruction for Neural Psychoacoustic Coding}

% It is OKAY to include author information, even for blind
% submissions: the style file will automatically remove it for you
% unless you've provided the [accepted] option to the icml2025
% package.

% List of affiliations: The first argument should be a (short)
% identifier you will use later to specify author affiliations
% Academic affiliations should list Department, University, City, Region, Country
% Industry affiliations should list Company, City, Region, Country

% You can specify symbols, otherwise they are numbered in order.
% Ideally, you should not use this facility. Affiliations will be numbered
% in order of appearance and this is the preferred way.
\begin{icmlauthorlist}
\icmlauthor{Dianwen Ng}{miro,ntu}
\icmlauthor{Kun Zhou}{tongyi}
\icmlauthor{Yi-Wen Chao}{ntu}
\icmlauthor{Zhiwei Xiong}{ntu}
\icmlauthor{Bin Ma}{tongyi}
\icmlauthor{Eng Siong Chng}{ntu}
\end{icmlauthorlist}

\icmlaffiliation{miro}{MiroMind, Singapore}
\icmlaffiliation{tongyi}{Tongyi Speech Lab, Alibaba Group, Singapore}
\icmlaffiliation{ntu}{College of Computing \& Data Science, Nanyang Technological University, Singapore}

\icmlcorrespondingauthor{Dianwen Ng}{dianwen001@e.ntu.edu.sg}

% You may provide any keywords that you
% find helpful for describing your paper; these are used to populate
% the "keywords" metadata in the PDF but will not be shown in the document
\icmlkeywords{Neural Audio Codec, Neural Psychoacoustic Codec, Zero-shot TTS, Vector Quantization}

\vskip 0.3in
]
\printAffiliationsAndNotice{}
% this must go after the closing bracket ] following \twocolumn[ ...

% This command actually creates the footnote in the first column
% listing the affiliations and the copyright notice.
% The command takes one argument, which is text to display at the start of the footnote.
% The \icmlEqualContribution command is standard text for equal contribution.
% Remove it (just {}) if you do not need this facility.

%\printAffiliationsAndNotice{}  % leave blank if no need to mention equal contribution
%\printAffiliationsAndNotice{\icmlEqualContribution} % otherwise use the standard text.

\begin{abstract}
Achieving high-fidelity audio compression while preserving perceptual quality across diverse audio types remains a significant challenge in Neural Audio Coding (NAC). This paper introduces MUFFIN, a fully convolutional Neural Psychoacoustic Coding (NPC) framework that leverages psychoacoustically guided multi-band frequency reconstruction. Central to MUFFIN is the Multi-Band Spectral Residual Vector Quantization (MBS-RVQ) mechanism, which quantizes latent speech across different frequency bands. This approach optimizes bitrate allocation and enhances fidelity based on psychoacoustic studies, achieving efficient compression with unique perceptual features that separate content from speaker attributes through distinct codebooks. MUFFIN integrates a transformer-inspired convolutional architecture with proposed modified snake activation functions to capture fine frequency details with greater precision. Extensive evaluations on diverse datasets (LibriTTS, IEMOCAP, GTZAN, BBC) demonstrate MUFFIN’s ability to consistently surpass existing performance in audio reconstruction across various domains. Notably, a high-compression variant achieves an impressive SOTA 12.5 Hz rate while preserving reconstruction quality. Furthermore, MUFFIN excels in downstream generative tasks, demonstrating its potential as a robust token representation for integration with large language models. These results establish MUFFIN as a groundbreaking advancement in NAC and as the first NPC system. Speech demos and codes are available \footnote{\url{https://demos46.github.io/muffin/}} \footnote{\url{https://github.com/dianwen-ng/MUFFIN}}.
\end{abstract}

\section{Introduction}

Neural Audio Coding (NAC) has emerged as a transformative technology in speech and audio processing, enabling high compression and high-fidelity reconstruction of audio signals with substantially reduced data representation sizes \cite{defossez2022high, du2024funcodec, wu2023audiodec, kumar2024high}. This advancement not only optimizes storage and transmission efficiencies but also enhances the integration of semantic and acoustic details into speech-large language models (LLMs) \cite{zhang2023speechgpt, zhang2024speechtokenizer, defossez2024moshi}. Within NAC, neural autoencoders and quantization techniques have achieved significant success in building unit representations that effectively capture essential audio features and compress them into precise, compact tokenized forms. Among these quantization methods, Residual Vector Quantization (RVQ) has emerged as the choice of method to enhance information preservation by refining the quantization process through multiple \textit{boosting turns} \cite{zeghidour2022soundstream}. This progressive refinement reduces reconstruction errors, enhancing fidelity to the original signal, ensuring higher quality audio outputs.

Psychoacoustics \cite{liu2017perceptually, zhen2020psychoacoustic, byun2022optimization}, the study of how humans perceive sound, underpins traditional audio coding frameworks such as MP3 and OPUS \citep{herre2019psychoacoustic}, yet remains largely unexplored in NAC. This discipline offers crucial insights for designing perceptually oriented systems, particularly through principles like perceptual masking. Different frequency bands carry distinct types of information: low frequencies are critical for speech intelligibility, mid frequencies capture formant structures essential for content articulation, and high frequencies convey speaker identity, pitch, and timbre—attributes integral to naturalness and spatial realism \cite{san2024discrete, petermann2023native}. By segmenting and encoding these bands separately, we account only for perceptually relevant distortions, reducing effective entropy while allowing greater controllability of learned speech attributes. This psychoacoustic perspective has the potential to redefine NAC, guiding more efficient compression strategies and richer, more robust reconstructions.

Leveraging psychoacoustic insights, we introduce MUFFIN, the first neural psychoacoustic codec (NPC) that utilizes the proposed multi-band spectral RVQ to achieve high-fidelity audio compression. MUFFIN aligns compression strategies with psychoacoustic principles, achieving an optimal balance between bitrate efficiency and perceptual quality. Additionally, we have enhanced frequency modeling through a novel modification of the snake activation function. Extensive experimental evaluations demonstrate MUFFIN’s superior performance in audio reconstruction compared to existing NACs. Furthermore, our codec excels in important downstream tasks, such as zero-shot text-to-speech (TTS), showcasing its potential as a robust token representations for integrating with LLMs for more advanced applications.

\textbf{Contributions} (1) We propose a novel multi-band spectral RVQ to optimize bitrate allocation, enhancing compression efficiency and perceptual audio quality. (2) We introduce MUFFIN, the \textbf{first neural psychoacoustic codec} that incorporates a modified snake activation function to achieve superior audio reconstruction at a SOTA 12.5 Hz compression while preserving high fidelity. (3) Our empirical demonstrations and experiments analyze the characteristics of each novel perceptual codebook, offering a better understanding of their potential use in various audio processing tasks.

\section{Related Work}
\textbf{Neural Audio Codec} Neural audio compression frameworks such as Soundstream \citep{zeghidour2022soundstream} and EnCodec \citep{defossez2022high} have advanced the field by implementing fully convolutional encoder-decoder architectures with RVQ bottlenecks, optimizing lossy tokenization while maintaining high fidelity. These models use HiFi-GAN vocoder losses \citep{kong2020hifi} to balance low bitrates with accurate reconstruction, ensuring robust audio quality even at reduced bandwidths \citep{ai2024apcodec}. AudioDec \citep{wu2023audiodec} further enhances this approach by integrating grouped convolutions for real-time processing in a streamable format, while HiFi-Codec \citep{yang2023hifi} introduces parallel group-RVQ layers to minimize redundancy. However, challenges persist. RVQ-based models often fail to preserve fine-grained details in complex acoustics, leading to perceptible artifacts \citep{langman2024spectral}. Transformer-based NAC models, like SpeechTokenizer \citep{zhang2023speechtokenizer}, which disentangle semantic content from the remaining audio features, face limitations in generalizing to input sequences longer than those encountered during training \citep{varivs2021sequence, chen2023once}, resulting in inconsistent performance and diminished output fidelity. Moreover, the prevalent practice of applying quantization across full frequency bands overlooks potential efficiency gains that could be achieved by accounting for the perceptual importance of different spectral regions, which is effective for achieving high compression capability while maintaining high-fidelity audio \citep{petermann2023native}. These issues highlight the need for continued innovation in neural audio coding to better address the complexities of auditory perception.

\textbf{Multi-band Audio Processing} 
Human auditory perception operates through a multi-band system, where the cochlea acts as a frequency analyzer, separating complex waveforms into bands that deliver crucial perceptual cues \citep{pulkki2015communication}. Psychoacoustic principles like critical bands and frequency masking are foundational to models such as the Bark scale \citep{zwicker1980analytical} and Equivalent Rectangular Bandwidth (ERB) \citep{moore1983suggested}, reflecting human auditory sensitivity to frequency variations. Leveraging these principles, multi-band modeling has significantly advanced audio compression by enhancing perceptual efficiency \citep{zhen2020psychoacoustic}. 

Legacy codecs like MPEG-1 Audio Layer III (MP3) and MPEG-2 Advanced Audio Coding (AAC) \citep{bosi1997iso} used these psychoacoustic models to dynamically allocate bits based on perceptual thresholds, achieving near-transparent sound quality. However, their static compression thresholds limited adaptability to complex audio signals. Modern neural audio codecs \citep{xiao2023multi,luo2024gull, langman2024spectral, nishimura2024hall, chen2024pyramidcodec} have overcome these limitations through flexible neural architectures that adapt more dynamically to signal complexities. For instance, the Penguins codec \citep{xiao2023multi} enhances perceptual audio quality by segmenting audio into frequency bands, using generative models for low frequencies and bandwidth extension for high frequencies to reduce bitrate without compromising quality. By quantizing mel-band features, spectral codecs \citep{langman2024spectral} align more closely with human auditory perception, although this can sometimes impact the accuracy of spectral details and phase information. Additionally, Gull \citep{luo2024gull} uses the Band-Split RNN (BSRNN) \citep{luo2023music} architecture to effectively capture temporal and inter-band dependencies through subband modeling and neural compression. However, its stacked BSRNN blocks and causal RNN layers increase computational latency, impacting real-time streaming applications.
%This design supports efficient audio modeling across various sample rates. However, despite these innovations, band-split techniques in neural audio codecs may increase computational burden which affects real-time streaming applications.

\section{Methods: Multi-Band Frequency Coding}
\subsection{Multi-band spectral residual vector quantization (MBS-RVQ)} 
In traditional methods, a multi-band model divides the frequency domain of a discrete-time signal, $x[n]$, into different frequency bands using the Fast Fourier Transform (FFT). Mathematically, the FFT of the signal can be expressed as:
\begin{equation}\resizebox{.88\linewidth}{!}{$
    {X}[k] = \sum_{n=0}^{N-1} x[n] e^{-j \frac{2\pi}{N} kn}, \quad k = 0, 1, \ldots, N-1$}
\end{equation}
where $X[k]$ represents the complex frequency components of the signal, $N$ is the total number of samples, and $k$ is the frequency index. We define $K$ non-overlapping frequency bands, each corresponding to a specific range of frequencies, denoted as $\mathbf{B}_k$ can be defined as:
\begin{equation}\resizebox{.85\linewidth}{!}{$
    \mathbf{B}_k = \{ f : f_{\min,k} \leq f < f_{\max,k} \}, \quad k = 1, 2,.., K$}
\end{equation}
where $f_{\min,k}$ and $f_{\max,k}$ denote the minimum and maximum frequencies of the $k^{th}$ band, respectively. After splitting the signal into frequency bands, we are able to construct the multi-band neural tokenizer by feeding each band with a typical time-domain neural autoencoder. The spectral matrix of each band is converted back into a time-domain waveform using the inverse FFT. This waveform is then passed through the encoder for quantization. Although this approach is straightforward for processing the signal in separate bands, it increases latency, as each band-pass signal requires individual encoding steps through the autoencoder (i.e., scaling the computational burden in terms of FLOPs by the number of $K$ bands) before being recombined to reconstruct the original signal. This added computational complexity limits its suitability for real-time streaming applications.

\begin{figure}[!ht]
    \centering
    \vspace{-0.2cm}
    \includegraphics[width=0.76\linewidth]{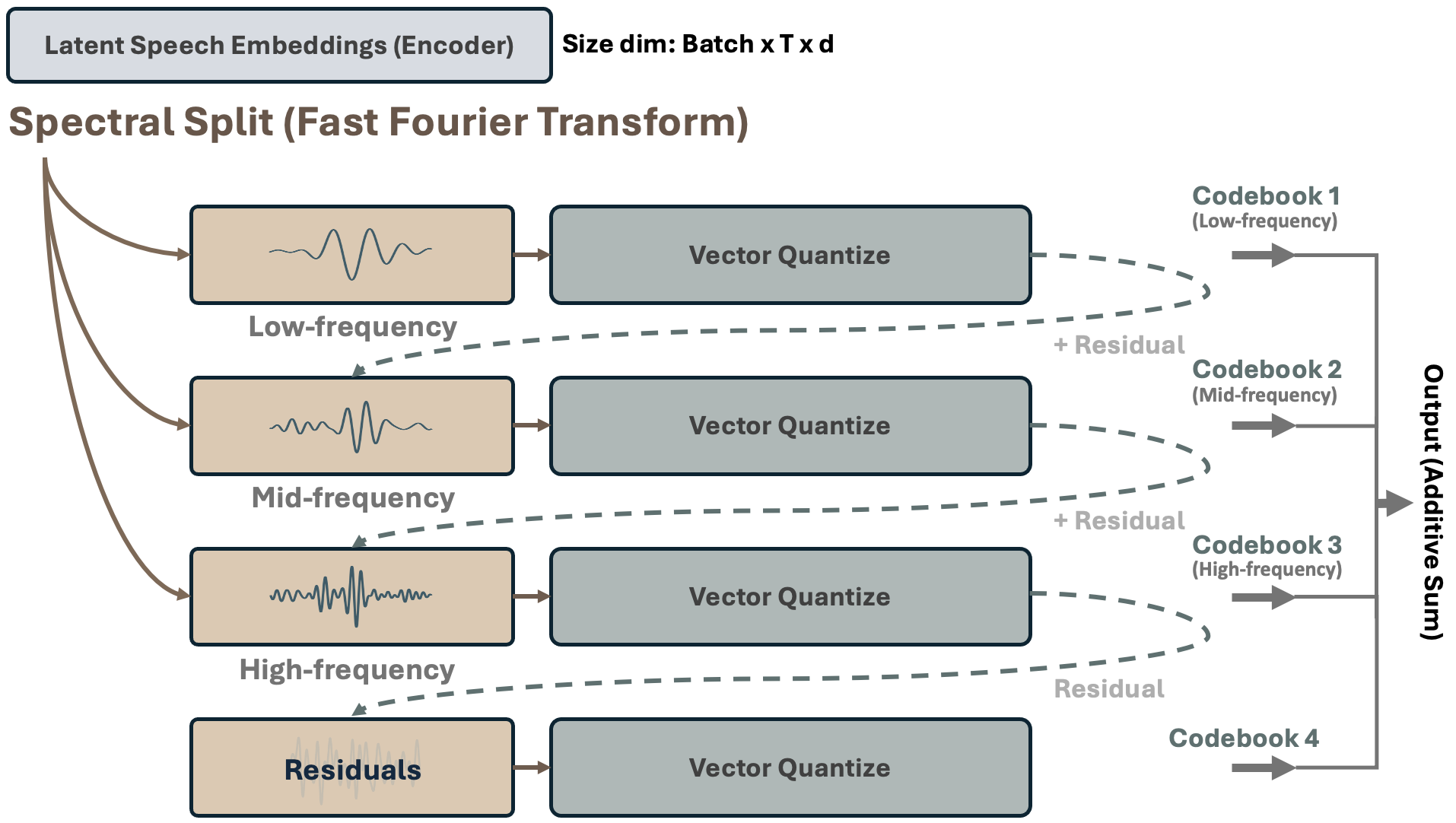}
    %\vspace{-0.3cm}
    \caption{Illustration of the MBS-RVQ process: Fast Fourier Transform (FFT) is applied to the encoded latent representation to isolate specific frequency bands, capturing targeted spectral information for each codebook. The filtered representation is reconstructed using inverse FFT before undergoing quantization. The quantization residuals are then passed to the next codebook 
    %as RVQ features, creating a hierarchical and progressively refined representation across codebooks.
    }
    \label{fig:rvq}
    \vspace{-0.3cm}
\end{figure}
Therefore, rather than splitting the bands at the input, we propose operating the multi-band processing within the encoder’s latent space using its latent features. The audio is first encoded into a compressed representation $z$, capturing both spectral and temporal features. The latent representation $z \in \mathbb{R}^{d \times T}$, where $d$ is the channel dimensionality and $T$ is the temporal length, is then decomposed into multiple frequency bands. Spectral band splitting is performed using the FFT, as previously detailed. We specifically target frequency bands of 0 to 18.75 Hz, 18.75 to 37.5 Hz, and 37.5 to 75 Hz, with corresponding scale factors of 4, 2, and 1, respectively. These scale factors are applied within the 75 Hz bandwidth of the latent representation, which is derived from compressing 24 kHz audio by a factor of 320 using the encoder. Each band is then quantized sequentially and continuously updated through an exponential moving average \citep{defossez2022high}. Note that the residuals from each subband quantizer will be accounted for in the subsequent subband feature input as residuals, following the RVQ approach, as we leverage the principle of successive refinement. This method enables each stage to quantize the residual error from the previous stage, thereby enhancing the preservation of detail in the audio representation. Figure \ref{fig:rvq} presents an illustration of the proposed MBS-RVQ.

Besides, adopting a multi-band splitting approach enhances the expression of individual unit codebooks within their respective spectral bands during the quantization process. This method is informed by psychoacoustic research, which highlights how different frequency ranges convey distinct types of information (Appendix \ref{appendix:a}). Specifically, low-frequency bands are essential for speech intelligibility due to their concentrated energy, while mid-frequency bands are crucial for articulating content through formant structures. High-frequency bands provide detailed acoustic features such as speaker identity, pitch, and timbre, contributing to the naturalness of speech \citep{san2024discrete, petermann2023native}. Furthermore, these higher frequencies enhance the spatial and ambient qualities of audio recordings.

By applying this knowledge, the encoding process can be optimized to allocate resources more effectively across the audio spectrum, ensuring that each data segment is processed to maximize the preservation and clarity of key auditory cues. This strategy not only allows for the perceptually meaningful quantization of audio units but also improves the efficiency and effectiveness of data compression, thereby significantly improving the overall perceptual quality of the output at a much lower codebook bitrate.

\subsection{Analysis: Multi-band modeling improves generative quality by leveraging the perceptual entropy bound.} \label{codes}
To provide justification for MUFFIN, let \(\mathbf{x}(t)\) be a continuous-time audio signal defined over \(t \in \mathbb{R}\), and let \(\hat{\mathbf{x}}(t)\) be its compressed approximation. Suppose the human auditory system is modeled by a set of perceptual filters that divide the frequency axis into \(K\) critical bands \(B_1, B_2, \dots, B_K\). Each band \(B_k\) corresponds to a region where the ear has distinct sensitivity levels (e.g., Bark or Mel scales). For each band \(B_k\), let \(P_k(\mathbf{x}(t))\) represent the perceptual threshold function indicating the maximum allowable distortion before artifacts become noticeable, and let \(D_k(\mathbf{x}(t), \hat{\mathbf{x}}(t))\) denote the band-specific distortion introduced by compression. The perceptual entropy \(E_p\) of \(\mathbf{x}(t)\) is defined as the minimal bit rate, $R_k$, needed so that \(D_k(\mathbf{x}(t), \hat{\mathbf{x}}(t)) \leq P_k(\mathbf{x}(t))\) for all \(k\), ensuring transparent audio compression: \[\resizebox{.88\linewidth}{!}{$
E_p = \sum_{k=1}^{K} \min \{ R_k : D_k(\mathbf{x}(t), \hat{\mathbf{x}}(t)) \leq P_k(\mathbf{x}(t)) \}.
$}\]
\begin{theorem}[Perceptual Entropy and Masking Bounds] \cite{cover1999elements}
\label{tm:perc}
For the given audio signal \(\mathbf{x}(t)\) and compressed representation \(\hat{\mathbf{x}}(t)\), the perceptual entropy \(E_p\) satisfies the following lower bound when optimal multiband modeling is employed:
\[\resizebox{.88\linewidth}{!}{$
E_p \;\geq\; \sum_{k=1}^{K} H(B_k \,\vert\, \mathbf{x}(t)) \;-\; \sum_{k=1}^{K} \Delta\!\bigl(B_k, \mathbf{x}(t)\bigr),$}
\]
where \(H(B_k \,\vert\, \mathbf{x}(t))\) is the Shannon entropy of the signal components within band \(B_k\), and \(\Delta(B_k, \mathbf{x}(t))\) is the perceptual masking effect that reduces the effective entropy by accounting for inaudible distortions in band \(B_k\).
\end{theorem}

Theorem \ref{tm:perc} characterizes how multiband modeling leverages psychoacoustic properties to achieve lower bit rates without sacrificing perceived audio quality. The first term, \(\sum_{k=1}^{K}H(B_k \mid \mathbf{x}(t))\), represents the total intrinsic entropy over the \(K\) critical bands, analogous to the information-theoretic bound one would calculate if there were no masking effects. However, human hearing does not require perfect fidelity in all frequency regions; many distortions remain hidden under the masking threshold \cite{zwicker2013psychoacoustics}. This phenomenon allows codecs to allocate fewer bits to masked regions without degrading the subjective audio experience. 

In mathematical terms, \(\Delta(B_k, \mathbf{x}(t))\) represents the ``masking offset" that effectively reduces the bits needed for band \(B_k\). Due to this offset, the total perceptual entropy \(E_p\) can be substantially smaller than the naive entropy sum, reflecting how frequency regions with strong masking require fewer bits for transparent encoding. This principle underlies the design of many perceptual audio codecs \cite{brandenburg1997overview}, where an FFT first decomposes the signal into subbands aligned with the human ear's critical bands, and then quantization is adapted based on psychoacoustic models \cite{johnston1988transform}.

\subsection{Model Architecture}
MUFFIN features an autoencoder architecture inspired by HiFi-Codec \citep{yang2023hifi}, utilizing a fully convolutional encoder-decoder network for temporal downscaling. We adopt the same striding configuration $(2, 4, 5, 8)$, optimized for 24kHz audio waveforms, achieving a total downsampling factor of 320 in the default configuration. A key component of our convolutional block is the multi-receptive field (MRF) fusion, adapted from \citet{kong2020hifi}, which aggregates outputs from residual blocks with varying dilated kernel sizes. This allows the model to capture dependencies across multiple temporal scales, enhancing its capacity to handle long-range sequential information.
\begin{figure}[!ht]
    \centering    \includegraphics[width=0.72\linewidth]{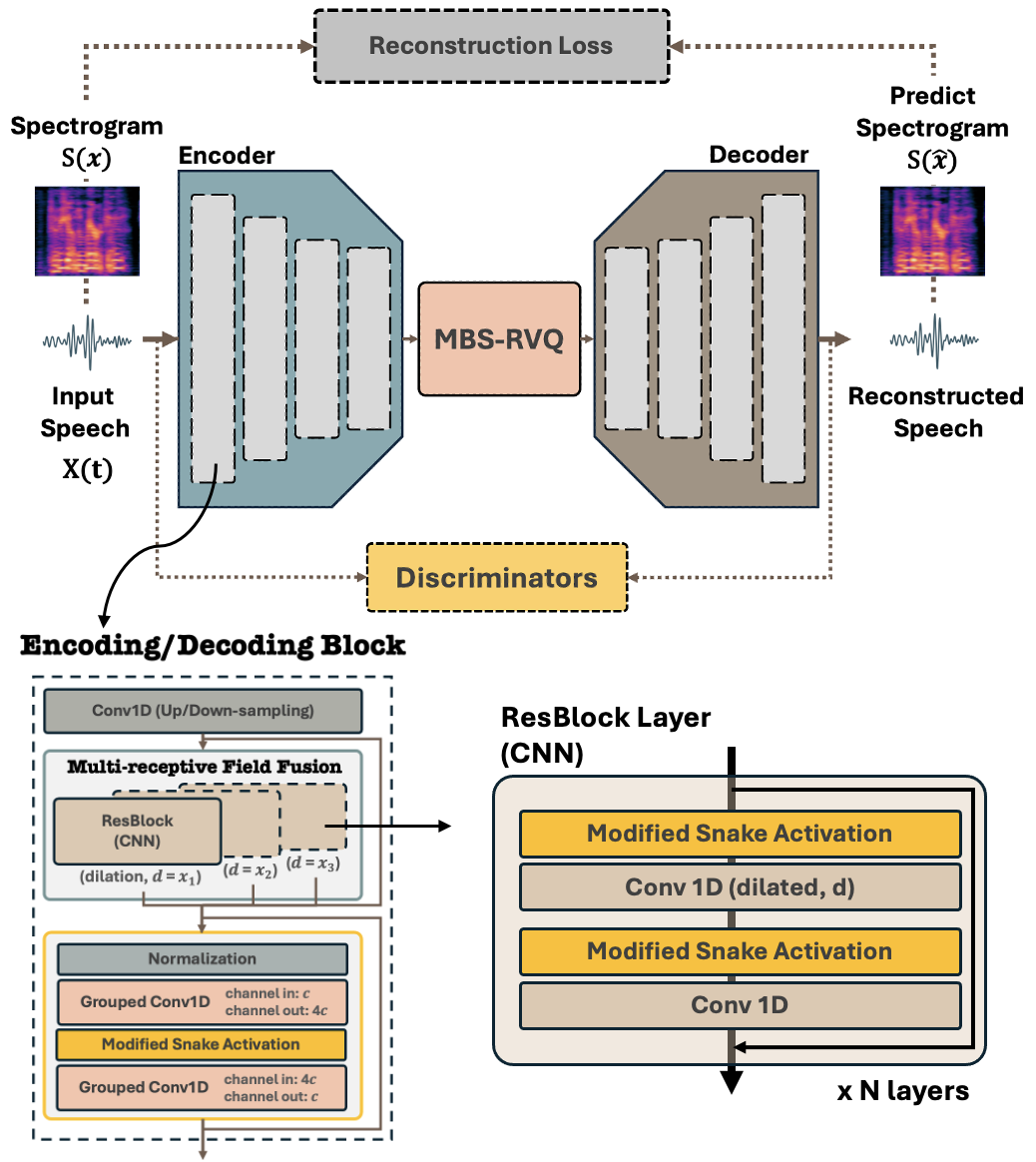}
    \vspace{-0.3cm}
    \caption{Architecture of MUFFIN incorporating a fully convolutional structure. 
    %The autoencoder blocks implement transformer-like operations through a (1) multi-receptive field communication layer for spatial dependency modeling, and (2) an inverted bottleneck layer for increased neural complexity. Besides, the layer block with a modified snake activation, as illustrated in the diagram, used to employ ReLU or LeakyReLU activations.
    }
    \label{fig:model}
    \vspace{-0.45cm}
\end{figure}
To enhance the model's representational power, we integrate an inverted bottleneck layer with residual skip connections, inspired by ConvNeXt \citep{liu2022convnet}, which increases channel dimensions and adds neural complexity. This design aligns with SOTA architectures \citep{dao2024transformers, yu2024mambaout, han2024demystify}, enabling the model to learn richer, more detailed representations, as shown in Figure \ref{fig:model}. To reduce latency by improving the computational efficiency from the channel-upsampling layer, we employ grouped convolutions that upscale in groups of 32 channels, which significantly reduces the number of model parameters and computations. The model consists of 46.1M parameters, with 34.2M in the encoder and 11.9M in the decoder. The Multiply-Accumulate Operations (MACs) reach 31.6G per second of audio sampled at 24 kHz, demonstrating enhanced performance efficiency compared to the HiFi-Codec's model, which has 61.5M parameters and taking 44.4G of MACs computational steps.

It is noteworthy that our convolutional block mirrors transformer functionality. The communication block from MRF mimics self-attention, enabling global temporal interactions, while the complexity block from the inverted bottleneck parallels the transformer’s feed-forward layer, adding depth and expressiveness. This transformer-inspired architecture efficiently captures hierarchical temporal patterns, ensuring robust audio signal reconstruction performance.

\textbf{Periodic Activation Function.} To enhance periodic modeling for better spectral preservation and high-fidelity reconstruction, we draw inspiration from \citet{kumar2024high} by replacing all Leaky ReLU activations with the snake activation function \citep{leebigvgan}, $x + 1 \text{/} \alpha \sin^2{\alpha x}$, which better preserves high-frequency information \citep{ziyin2020neural} and maintain Lipschitz continuity, since the derivative of snake activation function is bounded by a constant of 1. We further enhanced this function by introducing amplitude and bias adjustments to improve overall performance. In the original formulation, the parameter $\alpha$ controls the frequency of the periodic component, while its reciprocal $1 \text{/} \alpha$, attenuates the amplitude. To overcome this limitation, \citet{evans2024long} introduced a learnable scaling factor $\beta$ to adjust the amplitude independently. However, this adjustment may result in increased variance because of the amplified periodic magnitude relative to the input $x$ (Appendix \ref{appendix:d}, Figure \ref{fig:act}). To address this, we propose adding a bias term $\gamma$ that learns to shift the output, ensuring better adaptation to the new scale while preserving consistency with the input range. This approach mitigates variance issues and offers greater flexibility in fitting data, greatly enhancing high-fidelity reconstruction.

% In the course of modelling, Figure \ref{fig:act} shows four distinct data scenarios, where we observe that the vanilla activation function inadequately models regions exhibiting high-frequency patterns, likely due to trade-offs involving amplitude preservation. Although the introduction of the $\beta$ parameter mitigates this issue to some extent, it simultaneously introduces regions of elevated variance, which compromises overall stability. In contrast, the inclusion of a bias term $\gamma$ provides a more robust solution, effectively stabilizing the model and reducing the observed variance from over or underestimating outcome.

\subsection{Training Objectives}
\label{loss}
% MUFFIN is trained on typical reconstruction tasks to minimize the error in recreating audio signals, with the aim of producing outputs that are perceptually indistinguishable from the original audio. Our primary goal is to ensure that the reconstructed audio sounds natural and maintains high fidelity to the original input. This evaluation indirectly assesses the effectiveness of quantized speech in preserving information and capturing the expressivity of the input speech, which is crucial for high-fidelity reconstruction. 
Following the framework from HiFi-Codec, our training leverages two key components in the overall loss objectives.

\textbf{Reconstruction Loss} We employ a multiscale mel spectrogram reconstruction loss, calculated as the L1 distance between predicted and target mel-spectrograms over multiple time scales (i.e., a 64-bins mel-spectrogram derived from an STFT, with a window size of $2^i$ and a hop length of $2^i/4$ for $i = 7, 8, 9, 10, 11$). Unlike \citet{yang2023hifi} that uses 80 mel-spectrogram bins, we opted for a lower bin count based on perceptual evaluations, which revealed improved naturalness in the reconstructed audio while preserving key spectral features. Although reducing the bin count compromises frequency resolution, our results indicate that stricter distance reconstruction criteria, while preserving more information, may slightly harm perceptual quality.

\textbf{Discriminative Loss} We use three discriminators: a multi-scale STFT discriminator (MS-STFT) \citep{zeghidour2022soundstream}, a multi-period discriminator (MPD), and a multi-scale discriminator (MSD) \citep{kong2020hifi} to enhance perceptual quality through adversarial learning. We adopt the HingeGAN \citep{lim2017geometric} adversarial loss formulation and L1 feature matching loss \citep{kumar2019melgan}.

\textbf{MBS-RVQ Commitment Loss} We adopt quantization with commitment losses based on the VQ-VAE framework \citep{van2017neural}, using stop-gradients and the straight-through estimator \citep{bengio2013estimating} for backpropagation through the codebook lookup. The input is split into \(K\) critical frequency bands (\(B_i\)), where \(B_i\) denotes the \(i\)-th band. For each band, the encoder output \(z_e^{(B_i)}\) (encoded representation) is mapped to the nearest codebook vector \(z_q^{(B_i)}\) (quantized representation), ensuring frequency-specific representation. The commitment loss is defined as $\mathcal{L}_{\text{commit}} = \sum_{i=1}^K \beta_i \| z_e^{(B_i)} - z_q^{(B_i)} \|^2$, enabling precise mapping over the quantization of each spectral band.

%We employ the simple codebook and commitment losses with stop-gradients from the original VQ-VAE formulation \citep{van2017neural}, backpropagating gradients through the codebook lookup using the straight-through estimator \citep{bengio2013estimating}. Note that each codebook corresponds to a spectral band split, utilizing filtered information when learning to optimize the code searching process.

\section{Experiments}
\textbf{Data sources.} We train our model on a modest collection of 1,600 hours of speech, music, and environmental sounds. For speech, we use LibriTTS \citep{zen2019libritts} and EARS \citep{richter2024ears} datasets with expressive anechoic recordings of speech (585 and 100 hours, respectively). For music, we utilize Music4All \citep{santana2020music4all} (910 hours). For environmental sounds, we use ESC-50 \citep{piczak2015esc} (3 hours, 50 classes with 40 examples per class, loosely arranged into 5 major categories: animal, human, natural sounds, interior, and exterior sounds). Music and environmental sounds are used in MUFFIN to learn broader audio expression, enhancing the foundational representations. All audio was resampled to 24 kHz.

Prior to training, all audio files are truncated to a maximum duration of 10 seconds. A text file containing the paths to the processed audio files is provided to the data loader. This ensures a balanced sample distribution of speech and music file samples, mitigating data imbalance and preventing under-performance skewing towards vocal or instrumental reconstruction. During training, we apply voice activity detection to remove non-audio segments, optimizing learning efficiency. For each batch, 1-second audio segments are randomly selected from each instance and are zero-padded if shorter than 1 second. No additional data augmentation is applied to maintain simplicity in the experiment.

\textbf{Model and training details.} In our experiments, models were trained on two A800 GPUs for 300K iterations with a learning rate of 2e-4 and a batch size of 20 per GPU. All quantizers utilize a 9-bit code lookup from the EMA codebook. Additionally, we developed a low token-rate compression variant of MUFFIN to leverage the model's efficiency in capturing non-redundant information across different frequency bands. The highly compressed MUFFINs increase the downsampling rate by $960 \times$ and $1920 \times$ in the encoding layers, producing latent representations at 25 Hz and 12.5 Hz, respectively. This configuration achieves audio quantization at 150 and 100 tokens per second, utilizing additional residual codebooks with a total of six and eight vector quantization layers. The same MBS-RVQ configuration is used, maintaining a 4;2;1 frequency ratio relative to the sampled frequency of the latent embeddings (partitioned on a logarithmic scale) for the default band splits. MUFFIN, operating at 12.5 Hz, is trained on 2-second audio segments and requires four A800 GPUs to support a batch size of 10 per GPU. Using a lower learning rate is crucial to prevent gradient explosion. The high-compression MUFFIN model comprises approximately 50.6M parameters, with 36.5M in the encoder and 14.1M in the decoder (both models share the same architectural depth, differing only in downsampling rates). It achieves significantly lower MACs at 17.9G per second of audio, enabling faster inference.

Since our training setup closely adheres to Hifi-Codec, one of the leading codec model in the field, we establish it as our baseline by retraining the model with the same configuration, allowing for accurate performance comparison. Additionally, we benchmarked our method against other prominent codecs, including OPUS, Encodec and DAC, all reconfigured to a transmission rate of 3.0 kB/s to assess performance at similar transmission rate as our codec reconstruction model. Furthermore, we evaluate high-compression MUFFIN variant against Mimi, the recent SOTA codec model for 12.5 Hz, to highlight the strengths of our work.

\textbf{Evaluation objectives.} We utilize the objective evaluation metrics outlined in codec-SUPERB \citep{wu2024codec} to assess the perceptual quality of different audio domains. Specifically, we incorporate metrics such as the Perceptual Evaluation of Speech Quality (PESQ) \citep{rix2001perceptual}, Short-Time Objective Intelligibility (STOI) \citep{taal2010short}, STFT distance \citep{alsteris2007short}, Mel distance \citep{kubichek1993mel}, and F0CORR (F0 Pearson Correlation Coefficient) \citep{jadoul2018introducing}. The selection of these metrics for the corresponding audio domain is justified by the inclusion-exclusion criteria discussed in codec-SUPERB. Additionally, we employ two automated Mean Opinion Score (MOS) evaluation metrics, UTMOS \citep{saeki2022utmos} and ViSQOL \citep{hines2015visqol}, to assess the perceptual quality of the codecs. These metrics are designed to closely approximate subjective listening tests, providing a more accurate and robust evaluation of codec performance. For speech evaluation, we use the test-clean and test-other set from LibriTTS and evaluate emotional speech reconstruction with IEMOCAP \citep{busso2008iemocap}. For music, we employ the GTZAN dataset \citep{sturm2013gtzan}, while environmental sounds are evaluated using audio from the BBC sound effects \citep{laion_bbc_soundfx}.

\subsection{Experimental Results}
Table \ref{tab:libritts} and \ref{tab:iemocap} compare the speech reconstruction quality of our MUFFINs against existing top performing NACs. We indicate that NACs listed in the table operate at a bandwidth of 3.0 kB/s, except for Mimi at 1.0 kB/s, MUFFIN (default) at 2.7 kB/s, MUFFIN ($\triangledown$) at 1.35 kB/s, and MUFFIN ($\blacktriangle$) at 0.9 kB/s. We use LibriTTS evaluation dataset, with 4,837 samples for test-clean and 5,120 for test-other. Notably, MUFFIN achieves superior reconstruction fidelity, achieving the lowest distance error and outperforming competing NACs across various objective metrics. This includes HiFi-Codec, which serves as a robust baseline by being retrained under the same conditions, thereby confirming the effectiveness of the proposed framework. Furthermore, MUFFIN $\triangledown$ and $\blacktriangle$ achieve better UTMOS scores despite reductions in bandwidth and token rates. It appears that increasing the number of codebooks in our NAC system enhances the preservation of information from the original speech, resulting in improved perceptual quality as reflected by higher UTMOS scores. This suggests that with more codebooks, the system better captures nuanced details that contribute to the naturalness of the audio. However, while this strategy enhances perceptual naturalness, it may not uniformly improve all objective metrics. In fact, as the number of codebooks increases, especially at higher compression rates, some objective measures experience a decline. This decline may be attributed to the introduction of noise or artifacts by additional codebooks, which, while capturing more detail, also amplify aspects that negatively impact certain evaluation metrics. Thus, the relationship between increased codebooks and system performance exemplifies a trade-off between improved naturalness and the potential degradation of other audio quality metrics. Nevertheless, it is notable that MUFFIN $\blacktriangle$ remains competitive with leading NAC models and even outperforms Mimi in terms of naturalness and reconstruction fidelity, achieving significant gains in audio quality while maintaining efficient compression rates.
\begin{table}[!htb]
\centering \scriptsize
\tabcolsep=0.21cm
\renewcommand{\arraystretch}{1.3}
\vspace{-0.3cm}
\caption{Objective evaluation of reconstructed speech from the LibriTTS dataset using various neural audio codec models. \textit{GT} refers to the abbreviation for ground truth. Except for highly-compressed MUFFIN and Mimi, which uses a compression rate of $\triangledown: \times960$ (25.0 Hz) and $\blacktriangle: \times1920$ (12.5 Hz), the others have the compression rate of $\times320$ (75 Hz). Note that HiFi-Codec was retrained using the official configuration settings, but on the same dataset as MUFFIN.}
\vspace{0.1cm}
\begin{tabular}{lcccccc}
\hline
\multicolumn{7}{c}{Test-Clean (LibriTTS)}                                                                                             \\ \hline
\multicolumn{1}{l|}{Model}      & STFT           & MEL            & PESQ           & STOI           & UTMOS          & ViSQOL         \\ \hline
\multicolumn{1}{l|}{GT}         & -              & -              & -              & -              & 4.041          & -              \\
\multicolumn{1}{l|}{OPUS}       & 5.728          & 2.796          & 1.132          & 0.715          & 1.264          & 2.878          \\
\multicolumn{1}{l|}{Encodec}    & 1.956          & 1.051          & 2.042          & 0.903          & 2.269          & 4.078          \\
\multicolumn{1}{l|}{DAC}        & 1.759          & 0.849          & 2.370          & 0.915          & 2.951          & 4.143          \\
\multicolumn{1}{l|}{HiFi-Codec} & 1.618          & 0.765          & 2.712          & 0.943          & 3.831          & 4.410          \\
\multicolumn{1}{l|}{Mimi $\blacktriangle$}       & 2.488          & 1.706          & 1.715          & 0.620          & 2.966          & 3.791          \\
\multicolumn{1}{l|}{MUFFIN}     & \textbf{1.555} & \textbf{0.692} & \textbf{2.996} & \textbf{0.954} & 4.017          & \textbf{4.516} \\
\multicolumn{1}{l|}{MUFFIN $\triangledown$}     & 1.626          & 0.755          & 2.525          & 0.937          & 4.035          & 4.345          \\
\multicolumn{1}{l|}{MUFFIN $\blacktriangle$}     & 1.663          & 0.807          & 2.360          & 0.932          & \textbf{4.074} & 4.225          \\ \hline
\multicolumn{7}{c}{Test-Other (Libri-TTS)}                                                                                            \\ \hline
\multicolumn{1}{l|}{GT}         & -              & -              & -              & -              & 3.453          & -              \\
\multicolumn{1}{l|}{OPUS}       & 5.390          & 2.703          & 1.143          & 0.695          & 1.271          & 2.815          \\
\multicolumn{1}{l|}{Encodec}    & 1.998          & 1.119          & 1.960          & 0.888          & 2.026          & 4.017          \\
\multicolumn{1}{l|}{DAC}        & 1.813          & 0.913          & 2.220          & 0.897          & 2.497          & 4.053          \\
\multicolumn{1}{l|}{HiFi-Codec} & 1.681          & 0.840          & 2.419          & 0.919          & 3.216          & 4.296          \\
\multicolumn{1}{l|}{Mimi $\blacktriangle$}       & 2.515          & 1.688          & 1.611          & 0.612          & 2.498          & 3.679          \\
\multicolumn{1}{l|}{MUFFIN}     & \textbf{1.615} & \textbf{0.758} & \textbf{2.658} & \textbf{0.934} & 3.444          & \textbf{4.454} \\
\multicolumn{1}{l|}{MUFFIN $\triangledown$}     & 1.681          & 0.817          & 2.232          & 0.914          & 3.516          & 4.276          \\
\multicolumn{1}{l|}{MUFFIN $\blacktriangle$} &
  {1.725} &
  {0.875} &
  {2.086} &
  {0.904} &
  {\textbf{3.560}} &
  {4.129} \\ \hline
\end{tabular}
\label{tab:libritts}
\vspace{-0.4cm}
\end{table}
Table \ref{tab:iemocap} showcases MUFFIN’s zero-shot reconstruction capability on the full 12-hour IEMOCAP dataset, which contains expressive emotional speech that was not used during training.While reconstruction fidelity declines across the NACs, as shown in the objective metrics, both the default and high-compression variants of MUFFIN demonstrate superior robustness, achieving higher naturalness in human-perceived audio quality based on UTMOS scores compared to the ground truth references, despite the high compression. Nevertheless, the prominent drop in reconstruction fidelity for emotional content highlights a challenge in preserving emotional nuances, which could potentially impair emotional recognition in downstream tasks.
\begin{table}[!htb]
    \centering \scriptsize
    \tabcolsep=0.21cm
    \renewcommand{\arraystretch}{1.3}
    \vspace{-0.3cm}
    \caption{Objective evaluation of the reconstructed speech from the IEMOCAP dataset was conducted using various neural audio codec models, following the same setup as before.}
    \begin{tabular}{l|cccccc}
\hline
Model      & STFT  & MEL   & PESQ  & STOI  & UTMOS          & ViSQOL \\ \hline
GT         & -     & -     & -     & -     & 1.859          & -      \\
OPUS       & 2.361 & 1.586 & 1.207 & 0.478 & 1.242          & 2.642  \\
Encodec    & 2.150 & 1.290 & 1.649 & 0.746 & 1.321          & 3.501  \\
DAC        & 1.553 & 0.781 & 1.867 & 0.763 & 1.316          & 3.774  \\
HiFi-Codec & 1.447 & 0.755 & 1.998 & 0.763 & 1.564          & 3.651  \\
Mimi $\blacktriangle$      & 2.112 & 0.755 & 1.433 & 0.494 & 1.427          & 2.801  \\
MUFFIN & \textbf{1.399} & \textbf{0.675} & \textbf{2.178} & \textbf{0.806} & 1.903 & \textbf{4.000} \\
MUFFIN $\triangledown$    & 1.392 & 0.703 & 1.844 & 0.748 & \textbf{2.026} & 3.612  \\
MUFFIN $\blacktriangle$    & 1.429 & 0.754 & 1.726 & 0.723 & 2.019          & 3.376  \\ \hline
\end{tabular}
    \label{tab:iemocap}
\vspace{-0.2cm}
\end{table}
\begin{table}[!htb]
    \centering \scriptsize
    \tabcolsep=0.21cm
    \renewcommand{\arraystretch}{1.3}
    \vspace{-0.3cm}
    \caption{Objective evaluation of the reconstructed music from the GTZAN dataset using different NACs.}
\begin{tabular}{l|cccccc}
\hline
Model      & STFT  & MEL   & PESQ  & STOI  & F0CORR         & ViSQOL \\ \hline
OPUS       & 7.786 & 3.462 & 1.081 & 0.424 & 0.727          & 2.414  \\
Encodec    & 2.712 & 1.016 & 1.684 & 0.756 & 0.882          & 4.247  \\
DAC        & 2.493 & 0.928 & 1.709 & 0.741 & 0.867          & 4.220  \\
HiFi-Codec & 2.517 & 0.954 & 1.674 & 0.727 & \textbf{0.899}          & 4.170  \\
MUFFIN & \textbf{2.360} & \textbf{0.866} & \textbf{1.815} & \textbf{0.760} & 0.896 & \textbf{4.298} \\
MUFFIN $\triangledown$     & 2.492 & 0.928 & 1.474 & 0.674 & 0.879 & 4.273  \\
MUFFIN $\blacktriangle$    & 2.550 & 0.987 & 1.409 & 0.642 & 0.872          & 4.223  \\ \hline
\end{tabular}
\label{tab:gztan}
\vspace{-0.5cm}
\end{table}

Tables \ref{tab:gztan} and \ref{tab:bbc} present zero-shot reconstruction results on full data samples for music and audio, specifically from the GTZAN and BBC datasets. We observe a general decrease in fidelity for music reconstruction, as the task becomes more challenging due to the need to reconstruct multiple instrumental audio channels. From the table, we observe that HiFi-Codec achieves a higher F0CORR, indicating superior pitch accuracy and suggesting that its model structure better preserves vocal quality compared to other NACs. However, the difference in F0CORR between MUFFIN and HiFi-Codec is minimal, down to the finer decimal places, and MUFFIN consistently outperforms other NACs across the remaining metrics. Moreover, while MUFFIN $\triangledown\blacktriangle$ achieves a higher compression rate and learns to encode more efficiently with decent reconstruction fidelity for music, as confirmed by its competitive ViSQOL scores, we observed that PESQ and STOI were significantly lower, as reflected in Table \ref{tab:iemocap}. This suggests that highly compressed models face challenges in preserving fine information and are more vulnerable to reduced generalizability in zero-shot inference.

\begin{table}[!htb]
    \centering \scriptsize
    \tabcolsep=0.63cm
    \renewcommand{\arraystretch}{1.4}
    \vspace{-0.5cm}
    \caption{Objective evaluation of the reconstructed speech from the BBC dataset was conducted using various neural audio codec models, following the same setup as before.}
\begin{tabular}{l|ccc}
\hline
Model      & STFT           & MEL            & ViSQOL         \\ \hline
OPUS       & 6.093          & 2.984          & 1.000          \\
Encodec    & 1.998          & 1.011          & 3.852          \\
DAC        & 1.846          & 0.784          & 3.995          \\
HiFi-Codec & 1.773          & 0.795          & 4.009          \\
MUFFIN     & \textbf{1.658} & \textbf{0.720} & \textbf{4.065} \\
MUFFIN $\triangledown$     & 1.700          & 0.748          & 4.010          \\
MUFFIN $\blacktriangle$     & 1.706          & 0.777          & 3.997          \\ \hline
\end{tabular}
\label{tab:bbc}
\vspace{-0.4cm}
\end{table}

Table \ref{tab:bbc} showcases the strong generalizability of zero-shot reconstruction on general audio from the BBC dataset, underscoring the robustness and efficiency of MUFFINs when compared to other NACs. The results consistently demonstrate the improved quality of our neural codec for general audio reconstruction.

\subsection{Ablation Studies of Deconstructing MBS Codes}
\textbf{Auditory feature in codebook representations.} In this section, we examine the information encoded in MUFFIN's learned codebooks, which correspond to different auditory frequency bands. MUFFIN's codebooks focus on isolating the perceptual characteristics of speech attributes, guided by psychoacoustic research, and do so without the need for label-targeted supervision. To enhance understanding, we will also provide demos (Section F) of audio reconstructed from each codebook, accessible via the \href{https://demos46.github.io/muffin}{link}.

Codebook 1 (Low-frequency bands, 0 - 18.75 Hz) contains fundamental frequencies and strong harmonic content necessary for conveying core speech information. However, they primarily capture broad aspects of speech and lack detail in articulating speech content. This approach contrasts with previous NAC models where full-band RVQ often consolidates most speech information into the first codebook. To assess semantic content preservation in these low frequencies, we measure the STOI and word error rate (WER) using the pre-trained Whisper-large V3 model \citep{radford2023robust}, which analyzes audio reconstructed by the NAC model. For compatibility with the ASR model, trained on 16 kHz audio, we use the LibriSpeech test-clean dataset \citep{panayotov2015librispeech}, consisting of 2,620 samples that are resampled during reconstruction. Each sample’s speech decoded from individual codebooks is processed by the ASR model. We also evaluate other NAC models, including MUFFIN with vanilla RVQ, to explore how different codebooks affect the semantic information in speech.
\begin{table}[!htb]
    \centering \scriptsize
    \tabcolsep=0.09cm
    \renewcommand{\arraystretch}{1.3}
    \vspace{-0.5cm}
    \caption{The table presents the WER of ASR performance on reconstructed speech from each NAC’s codebook using the Whisper-large V3 pre-trained model.}
\begin{tabular}{lcccccccccc}
\hline
 & \multicolumn{2}{c}{MUFFIN} & \multicolumn{2}{c}{RVQ} & \multicolumn{2}{c}{Hifi-Codec} & \multicolumn{2}{c}{DAC} & \multicolumn{2}{c}{Encodec} \\ \hline
GT         & \multicolumn{10}{c}{WER: 2.41; STOI: -}                                  \\ \hline
           & WER  & STOI  & WER  & STOI  & WER  & STOI  & WER  & STOI  & WER  & STOI  \\ \hline
All  & \textbf{2.67} & 0.940 & 2.72 & 0.930 & 3.00 & 0.919 & 3.53 & 0.901 & 3.15 & 0.900 \\ \hline
Code 1 & 70.3 & 0.644 & 75.6 & 0.702 & 154  & 0.572 & 36.1 & 0.731 & 33.7 & 0.764 \\
Code 2 & 114  & 0.379 & 141  & 0.426 & 139  & 0.454 & 132  & 0.148 & 159  & 0.199 \\
Code 3 & 191  & 0.436 & 100  & 0.157 & 100  & 0.090 & 100  & 0.079 & 153  & 0.121 \\
Code 4 & 107  & 0.082 & 101  & 0.086 & 112  & 0.129 & 100  & 0.049 & 147  & 0.094 \\ \hline
\end{tabular}
\label{tab:cb1}
\vspace{-0.5cm}
\end{table}

Table \ref{tab:cb1} highlights the crucial role of Codebook 1 in the MUFFIN model, as it becomes significantly challenging to recognize content using any codebook other than Codebook 1, with recognition errors exceeding 100. Unlike vanilla RVQ, which typically consolidates most speech content into the first codebook, MUFFIN strategically distributes information between Codebooks 1 and 2. The main speech intelligibility is allocated to Codebook 1, while details related to articulation are captured in the mid-frequency range of Codebook 2, which provides it with more contextual information than Codebook 3. Notably, MUFFIN’s Codebooks 1 and 2 together achieve an STOI of 0.729 and WER of 19.2. In contrast, NACs that employ traditional RVQ often show a progressive decline in semantic content across subsequent codebooks, accompanied by a noticeable decrease in STOI.

Codebook 3 (High-frequency bands, 37.5 - 75 Hz) captures essential auditory cues, such as speaker identity, pitch, and timbre, which are crucial for distinguishing speakers and enriching the depth of reconstructed audio. To assess the effectiveness of each codebook in preserving speaker information, we randomly select 600 speech files from the VoxCeleb dataset \citep{nagrani2017voxceleb}, each representing one of six distinct speakers. Latent features are extracted from each codebook and then average-pooled to form a vector representation for each speech sample. Using t-SNE \citep{van2008visualizing}, we visualize these vector representations in a two-dimensional space to identify potential clusters corresponding to different speakers, as presented in Figure \ref{fig:ch5tsne} (Appendix \ref{appendix:e}, Figure \ref{fig:ch5tsne}).

% \begin{figure}[!ht]
%     \centering    \includegraphics[width=0.75\linewidth]{figures/tsne.png}
%     \caption{A t-SNE plot showcasing each codebook, with speech randomly sampled from VoxCeleb, effectively represents six distinct speakers of the color code.}
%     \label{fig:ch5tsne}
% \end{figure}

From Figure \ref{fig:ch5tsne}, \textbf{Codebook 1} exhibits a broad, dispersed distribution with some cluster overlap, suggesting it encodes foundational speech content with substantial variance across samples. \textbf{Codebook 2} forms a more concentrated central cluster with significant overlap between speakers, implying shared features likely related to vowel and consonant articulation. \textbf{Codebook 3}, in contrast, shows well-separated clusters with the most distinct boundaries among speakers and minimal overlap, indicating its focus on capturing speaker-specific features. This distinct clustering should lead to a low distance error in speaker group classification, suggesting that Codebook 3 is highly effective at distinguishing between different speakers based on their unique vocal attributes. Additionally, we strongly suggest comparing the auditory results between Codebook 3 and others using the provided demos to gain a more intuitive understanding. Lastly, \textbf{Codebook 4} presents a compact distribution with considerable class overlap, suggesting it primarily encodes random or residual features that contribute minimally to core speech information. Additionally, the variations and the distributive patterns in the t-SNE plots suggest that each codebook captures slightly different information. 
%and likely serves a distinct purpose.

\subsection{Zero-shot TTS with MUFFIN tokens}
We further evaluate the performance of different codec systems in a zero-shot text-to-speech (TTS) setting. For this evaluation, we use VALL-E \cite{wang2023neural} as the TTS framework, implementing it based on a popular release\footnote{\url{https://github.com/lifeiteng/vall-e}}, and trained it on LibriTTS dataset. VALL-E employ a decoder-only language model, predicting the first layer of acoustic codes autoregressively, while the remaining layers are predicted non-autoregressively.  It has gained popularity as a zero-shot TTS framework and is frequently used as a benchmark in various studies \cite{du2024cosyvoice, shen2023naturalspeech2}.

We follow the experimental settings in \cite{zhou2024phonetic} and test 3 different VALL-E systems on the test-clean set from LibriTTS. For each speaker in the test set, one speech sample from the same speaker is randomly selected as the prompt, ensuring it is distinct from the speech to be synthesized. All the synthesized speech are ranging from 3 to 10 seconds. We first calculate the word error rate (WER) of the synthesized speech, as shown in Table \ref{tab:tts}. Compared to VALL-E using Encodec or Hifi-Codec, VALL-E with MUFFIN achieves the lowest WER results, demonstrating superior robustness in the synthesized speech. We then report the subjective evaluation results in Table \ref{tab:tts}, where 10 native speakers evaluated a total of 80 synthesized speech samples. Each speech sample was rated on a 5-point scale for speech quality and naturalness (MOS), as well as speaker similarity to the prompt speech (S-MOS) . Consistent with the objective results, the VALL-E system using MUFFIN achieves the highest scores for both speech quality and speaker similarity. These findings suggest that MUFFIN effectively disentangles between semantic and acoustic information, thereby reducing error propagation during the TTS training and resulting in improved speech quality during inference. 

\begin{table}[t]
\centering \scriptsize
    \tabcolsep=0.15cm
    \vspace{-0.3cm}
    \renewcommand{\arraystretch}{1.5}
    \caption{The zero-shot TTS results include word error rate (WER), mean opinion score (MOS) for speech quality and naturalness, and speaker mean opinion score (S-MOS) for speaker similarity, evaluated for VALL-E with three different codecs. MOS and S-MOS are reported with 95 \% confidence interval.}
\begin{tabular}{l|cccc}
\hline
Systems             & WER (\%) & MOS  & S-MOS   &  SECS \\ \hline
VALL-E w/ Encodec   & 21.05    & 3.91 $\pm$ 0.287    &  3.70 $\pm$ 0.368   &  0.5914   \\ 
VALL-E w/ HiFi-Codec & 32.35    &  4.00 $\pm$ 0.532   &  4.04 $\pm$ 0.333  &  0.5874   \\ 
VALL-E w/ MUFFIN    &  \textbf{12.20}   &  \textbf{4.18 $\pm$ 0.278}   &   \textbf{4.19 $\pm$ 0.288}  &  \textbf{0.6099}    \\ \hline
\end{tabular}
\label{tab:tts}
\vspace{-0.7cm}
\end{table}

\section{Conclusion}
In conclusion, we introduced MUFFIN, a neural psychoaudio codec that offers a novel perspective on quantizing units with the proposed MBS-RVQ within the latent space. By strategically aligning the codec architecture with psychoacoustic principles, MUFFIN achieves an optimal balance between compression efficiency and perceptual fidelity, tackling longstanding challenges in the domain. Extensive evaluations demonstrate MUFFIN’s SOTA performance across diverse audio types reconstruction and downstream zero-shot TTS task. Furthermore, the development of a 1920 times highly compressed MUFFIN variant underscores its ability to maintain perceptual quality even under extreme compression settings. This study lays the groundwork for future advancements in real-time low latency neural audio coding and its integration with LLMs, providing a robust and scalable solution for speech-LLMs model applications.

% Acknowledgements should only appear in the accepted version.
\section*{Acknowledgements}
This research was conducted at the Alibaba-NTU Singapore Joint Research Institute.

\section*{Impact Statement}
The development of MUFFIN, a high-fidelity Neural Psychoacoustic Coding (NPC) framework, represents a novel and effective alternative to conventional speech encoders \cite{hsu2021hubert, chen2022wavlm, ng2023hubert}. This framework offers broad applicability across domains including media streaming, telecommunications, and assistive technologies. By incorporating psychoacoustic principles to guide perceptual compression, MUFFIN enables the transmission of high-quality audio at significantly lower bitrates. This advancement facilitates wider accessibility in bandwidth-constrained environments, thereby promoting digital inclusion and improving access to communication and entertainment services in regions with limited internet infrastructure. In addition to its compression efficiency, MUFFIN introduces a mechanism for disentangling speaker identity from speech content. This capability offers new opportunities for personalization and content manipulation but also raises critical ethical considerations related to privacy, data security, and the potential for misuse in synthetic or manipulated speech, such as deepfake audio. These concerns highlight the urgent need for responsible deployment practices and appropriate regulatory oversight. Moreover, the integration of MUFFIN with large-scale language models has the potential to significantly enhance human-computer interaction in areas such as creativity support, education, and accessibility-driven applications. This synergy positions MUFFIN as a foundational component in the next generation of AI-powered multimedia systems. Future research should focus on developing robust bias mitigation strategies and establishing clear ethical frameworks to guide the deployment of neural audio coding technologies in a fair and secure manner across diverse global populations.
%Authors are \textbf{required} to include a statement of the potential broader impact of their work, including its ethical aspects and future societal consequences. This statement should be in an unnumbered section at the end of the paper (co-located with Acknowledgements -- the two may appear in either order, but both must be before References), and does not count toward the paper page limit. In many cases, where the ethical impacts and expected societal implications are those that are well established when advancing the field of Machine Learning, substantial discussion is not required, and a simple statement such as the following will suffice:``This paper presents work whose goal is to advance the field of Machine Learning. There are many potential societal consequences of our work, none which we feel must be specifically highlighted here.''The above statement can be used verbatim in such cases, but we encourage authors to think about whether there is content which does warrant further discussion, as this statement will be apparent if the paper is later flagged for ethics review.

% In the unusual situation where you want a paper to appear in the
% references without citing it in the main text, use \nocite
%\nocite{langley00}

\bibliography{example_paper}
\bibliographystyle{icml2025}

%%%%%%%%%%%%%%%%%%%%%%%%%%%%%%%%%%%%%%%%%%%%%%%%%%%%%%%%%%%%%%%%%%%%%%%%%%%%%%%
%%%%%%%%%%%%%%%%%%%%%%%%%%%%%%%%%%%%%%%%%%%%%%%%%%%%%%%%%%%%%%%%%%%%%%%%%%%%%%%
% APPENDIX
%%%%%%%%%%%%%%%%%%%%%%%%%%%%%%%%%%%%%%%%%%%%%%%%%%%%%%%%%%%%%%%%%%%%%%%%%%%%%%%
%%%%%%%%%%%%%%%%%%%%%%%%%%%%%%%%%%%%%%%%%%%%%%%%%%%%%%%%%%%%%%%%%%%%%%%%%%%%%%%
\newpage
\appendix
\onecolumn
\section{Psychoacoustic evidence of perceptual speech characteristics}
\label{appendix:a}
\begin{figure}[!ht]
\centering\includegraphics[width=0.95\linewidth]{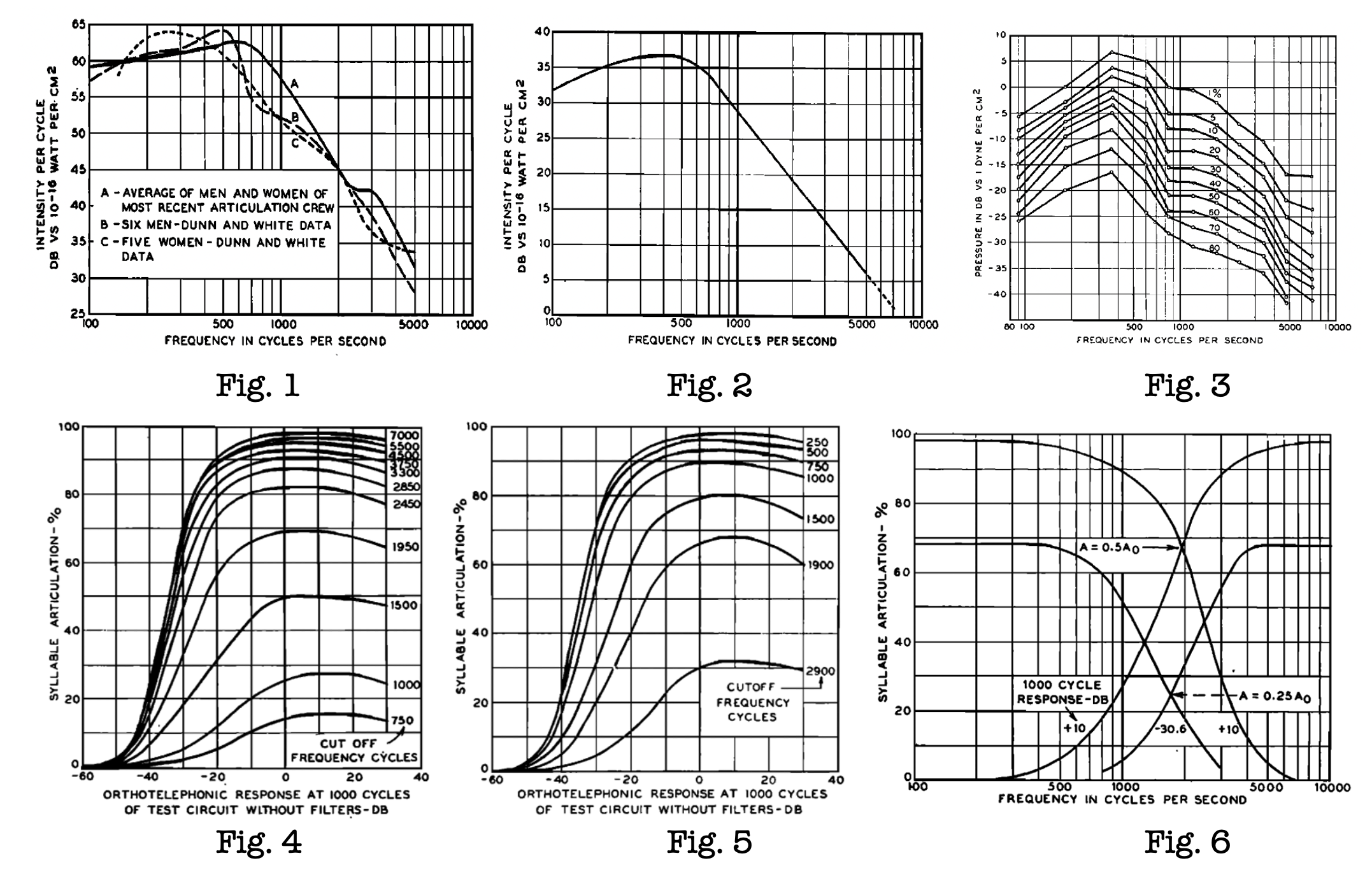}
    \caption{The figures have been sourced from \citet{french1947factors}, which discusses how speech sounds are recognized by the ear. The data were collected from microphones based on human speech and then analyzed with computational tools to derive the intensity and sound pressure levels. (1) Comparison of Speech Spectra. (2) Idealized Long Average Speech Spectrum at one meter from lips. (3) R.m.s. pressure of speech at 30cm from lips. (4) Articulation test with low pass filters. (5) Articulation test with high pass filters. (6) Syllable articulation versus cut-off frequency.}
    \label{fig:psy}
\end{figure}

From the presented illustration, it becomes clear that the core components of speech intelligibility are predominantly emphasized within the lower frequency bands where the majority of vocal energy is concentrated. Fig. 1 demonstrates the variation in intensity levels across the frequency spectrum, measured in decibels relative to a standard sound pressure level. This figure contrasts intensity levels at a near-field distance, merely two inches from the speaker's lips, with adjustments made for gender differences in voice power and frequency content. The data reveal a pronounced convex curve with higher intensities noted in the lower frequencies, which gradually decrease as the frequency increases. This pattern supports our initial assertions and is corroborated by Fig. 2 and 3. These figures extend the observations to longer ranges (1 meter) and include measurements of RMS pressure, following the experimental setup described by \citet{dunn1940statistical} for six male subjects.

Then, Fig. 4 and 5 present the results from articulation tests employing low-pass and high-pass filters across varying cutoff frequencies, demonstrating the impact of frequency range restrictions on speech intelligibility. Notably, a reduction in the cutoff frequency through low-pass filters correlates with a decrease in intelligibility, highlighting the critical role of higher frequencies in the recognition of consonants and the differentiation of similar-sounding syllables. In contrast, high frequencies are pivotal for capturing the intricate details and nuances that enhance speech clarity and comprehensibility.

Interestingly, although high-pass filtering up to a specific cutoff frequency can yield improvements in speech intelligibility, performance degrades when lower frequencies are excessively attenuated. This trend underscores the critical role of low-frequency components—primarily conveyed through vowel sounds—in preserving the spectral power and tonal richness of speech. These vowel sounds are fundamental to intelligibility, as they provide essential acoustic energy and rhythmic structure.

Fig. 6 integrates the findings from Fig. 4 and 5, combining the effects of high-pass and low-pass filtering to offer a more holistic view of how frequency components influence speech intelligibility across various system settings. The intersection of the filter curves in Fig. 6 identifies critical bands, pinpointing a crucial frequency range central to maintaining speech intelligibility. This analysis elucidates that speech articulation predominantly occurs within the mid-frequency bandwidth, providing key insights into the frequency-dependent nature of speech processing.

\section{Validity of MBS-RVQ at latent representation space.} 
We note the theoretical distinction in performing band splitting at the input space versus applying it to the latent representations of an autoencoder. This raises key considerations regarding the preservation of psychoacoustic properties within the latent representation space. However, we argue that latent representations retain the psychoacoustic properties of speech is supported by the Lipschitz continuity of the encoder. 

Formally, if \(f: \mathbb{R}^n \to \mathbb{R}^m\) is Lipschitz continuous with constant \(L\), then for any two speech signals \(\mathbf{x}_1,\mathbf{x}_2\) \cite{hager1979lipschitz}:
\begin{equation}
    \| f(\mathbf{x}_1) - f(\mathbf{x}_2) \| \;\le\; L \,\|\mathbf{x}_1 - \mathbf{x}_2\|
\end{equation}
Psychoacoustic cues such as formant positions, harmonic relationships, and energy distributions in critical frequency bands are primarily reflected in subtle variations of the speech signal's waveform. Lipschitz continuity ensures that these modest yet perceptually crucial differences are neither excessively magnified nor erased when the signal is transformed into the latent domain. In other words, two psychoacoustically similar speech signals cannot become drastically separated in latent space \cite{arjovsky2017wasserstein, bartlett2002rademacher}. Furthermore, empirical evidence in representation learning supports this notion, demonstrating that neural networks constrained by Lipschitz continuity typically learn more structured representations, within which subtle perceptual attributes remain discernible \cite{belkin2003laplacian, gulrajani2017improved}.

Consequently, if an autoencoder's encoder maintains Lipschitz continuity, the latent embeddings it generates for speech signals can be expected to closely reflect the psychoacoustic characteristics present in the original waveform. Minor spectral changes perceived by listeners, such as slight shifts in vowels or sibilants, correspond to small changes in latent space, thus helping to preserve the overall psychoacoustic signature of the speech. This argument provides the basis for the notion that, although a strict one-to-one psychoacoustic fidelity is not mathematically guaranteed, in practice, Lipschitz continuity significantly mitigates the risk of losing important auditory details in the encoder’s output.

In this work, we introduce the MUFFIN encoder, a novel architecture that is provably Lipschitz continuous. The core design leverages primarily linear components—namely convolutional layers and fully connected (linear) layers—alongside a modified Snake activation function. Convolutional and linear layers are intrinsically linear transformations, ensuring that the overall network adheres to Lipschitz continuity. 

For modified snake activation function, \( f(x) = x + \frac{\beta}{\alpha}\sin^2(\alpha x) + \gamma \),  the derivative is presented as:

\[
f'(x) = \frac{d}{dx}\bigl[ x + \frac{\beta}{\alpha}\sin^2(\alpha x) + \gamma\bigr] 
       = 1 + 2\beta\,\sin(\alpha x)\cos(\alpha x)
       = 1 + \beta \sin(2\alpha x).
\]

Since \(\sin(2 \alpha x)\) is bounded between \(-1\) and \(1\), we have

\[
|f'(x)| = |\beta \,\sin(2 \alpha x)| \leq |\beta|.
\]

A function whose derivative is bounded by \(L\) is \(L\)-Lipschitz. Therefore,

\[
|f'(x)| \leq |\beta| \quad \Longrightarrow \quad \text{\(f\) is \(|\beta|\)-Lipschitz.}
\]

Hence, our modified snake activation function is Lipschitz continuous with a Lipschitz constant \(|\beta|\). This property ensures stability in our model, as the layer computations with the activation functions preserve the distances between input features.

\section{On the effect of MBS-RVQ compared to vanilla RVQ.}

In this section, we conduct an ablation study by comparing the performance with disabling multi-band spectral residual vector quantization, reverting to vanilla residual vector quantization, to assess the contribution of MBS-RVQ in enhancing generative quality under the constraint imposed by perceptual entropy. In addition to the results presented in Table \ref{tab:cb1}, which analyze the individual contributions of each codebook and the full MBS-RVQ quantizer configuration with respect to word error rate and short-time objective intelligibility, we further report detailed reconstruction performance of the neural psychoacoustic codec. The following tables present reconstruction quality evaluated using the same metrics employed in the main results. Complementary to the table-based evaluation, we also provide audio samples on our demo page to audibly highlight the differences between MBS-RVQ and vanilla RVQ.

\begin{table}[!htb]
    \centering \footnotesize
    \tabcolsep=0.35cm
    \renewcommand{\arraystretch}{1.3}
    \caption{The table presents the objective evaluation of the reconstructed speech from listed evaluation set using full codebook quantizers of MBS-RVQ versus vanilla RVQ.}
    \vspace{0.1cm}
\begin{tabular}{lcccccc}
\hline
\multicolumn{7}{l}{Test-Clean (LibriTTS)}                                                   \\ \hline
\multicolumn{1}{l|}{Model}                 & STFT  & MEL   & PESQ  & STOI  & UTMOS & ViSQOL \\ \hline
\multicolumn{1}{l|}{MUFFIN}                & 1.555 & 0.692 & 2.996 & 0.954 & 4.017 & 4.516  \\
\multicolumn{1}{l|}{Vanilla-RVQ}           & 1.627 & 0.768 & 2.856 & 0.940 & 3.875 & 4.328  \\
\multicolumn{1}{l|}{MUFFIN (12.5 Hz)}      & 1.663 & 0.807 & 2.360 & 0.932 & 4.074 & 4.225  \\
\multicolumn{1}{l|}{Vanilla-RVQ (12.5 Hz)} & 1.755 & 0.879 & 2.260 & 0.924 & 3.785 & 4.017  \\ \hline
\multicolumn{7}{l}{Test-Other (LibriTTS)}                                                   \\ \hline
\multicolumn{1}{l|}{MUFFIN}                & 1.615 & 0.758 & 2.658 & 0.934 & 3.444 & 4.454  \\
\multicolumn{1}{l|}{RVQ}                   & 1.683 & 0.810 & 2.544 & 0.917 & 3.318 & 4.268  \\
\multicolumn{1}{l|}{MUFFIN (12.5 Hz)}      & 1.725 & 0.875 & 2.086 & 0.904 & 3.560 & 4.129  \\
\multicolumn{1}{l|}{RVQ (12.5 Hz)}         & 1.863 & 0.963 & 1.940 & 0.815 & 3.399 & 3.993  \\ \hline
\multicolumn{7}{l}{IEMOCAP}                                                                 \\ \hline
\multicolumn{1}{l|}{MUFFIN}                & 1.399 & 0.675 & 2.178 & 0.806 & 1.903 & 4.000  \\
\multicolumn{1}{l|}{RVQ}                   & 1.510 & 0.793 & 2.039 & 0.715 & 1.805 & 3.883  \\
\multicolumn{1}{l|}{MUFFIN (12.5 Hz)}      & 1.429 & 0.754 & 1.726 & 0.723 & 2.026 & 3.612  \\
\multicolumn{1}{l|}{RVQ (12.5 Hz)}         & 1.584 & 0.835 & 1.644 & 0.645 & 1.917 & 3.455  \\ \hline
\end{tabular}
\end{table}

\section{Periodic Activation Function -- Modified Snake Activation Function}
\label{appendix:d}
\begin{figure}[!ht]
\centering\includegraphics[width=0.8\linewidth]{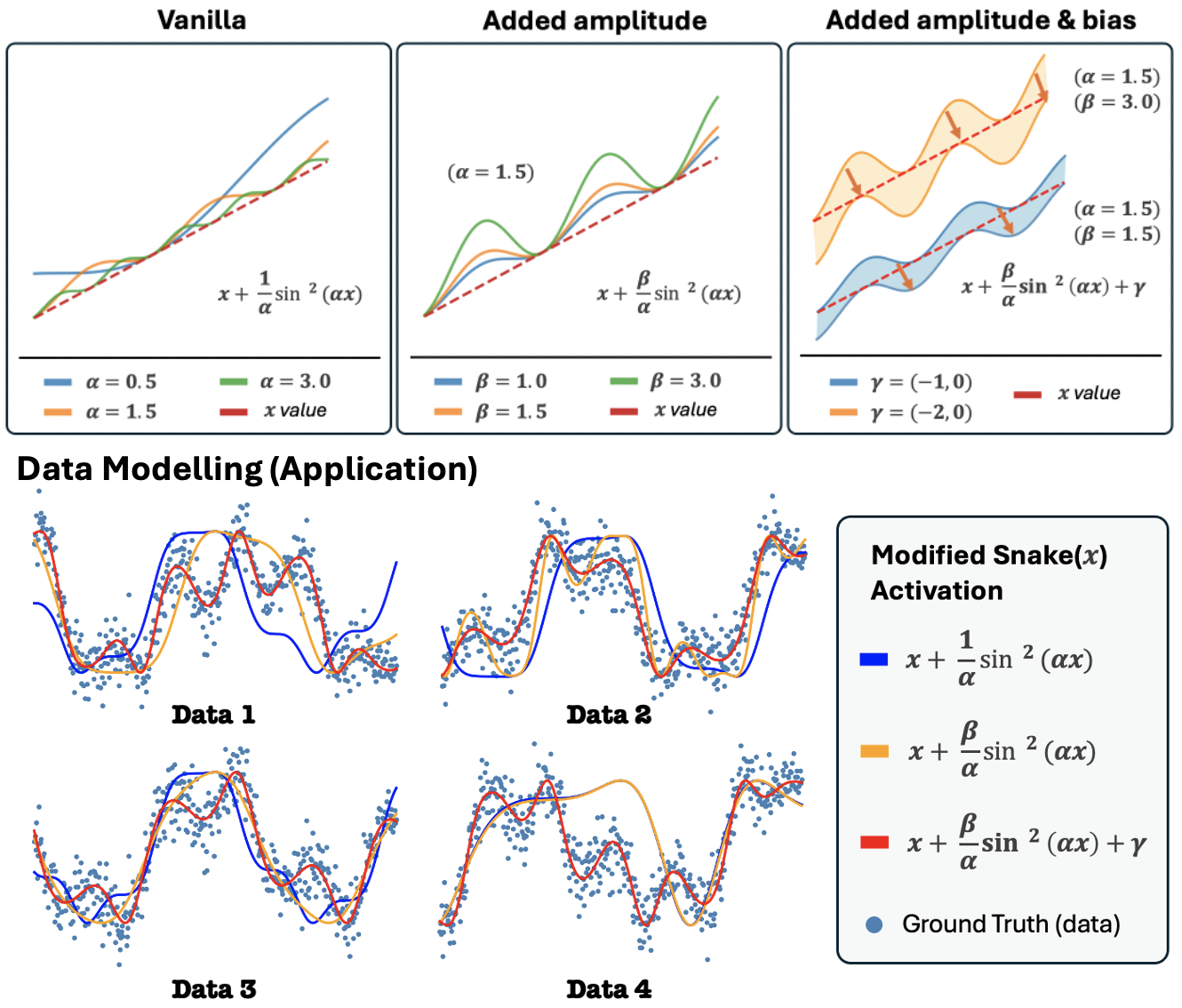}
    \caption{Illustration of our proposed modifications to the vanilla snake activation and its behavior in actual modeling for different sequential data.}
    \label{fig:act}
\end{figure}

In the course of modeling, Figure \ref{fig:act} depicts four distinct data scenarios, each presenting challenging frequency details where periodic patterns appear odd and highly abrupt, complicating the modelling. It is observed that the standard activation function inadequately models regions exhibiting high-frequency patterns, likely due to trade-offs involving amplitude preservation. Although the introduction of the amplitude, \(\beta\) parameter, partially mitigates this issue, it simultaneously introduces regions of elevated variance, which compromise overall stability. In contrast, we observe that incorporating a bias term \(\gamma\) provides a more robust solution by effectively stabilizing the model and reducing the variance associated with overestimating or underestimating outcomes.

In addition to the figure above, we present an ablation study examining the impact of each additive term on reconstruction quality in Table \ref{tab:act_fn}. These results are evaluated using the same metrics as those employed in the main table and the table elucidates the distinct contributions of each additive term to the reconstruction quality of the codec (i.e., the component with amplitude and bias).

\begin{table}[!htb]
    \centering \footnotesize
    \tabcolsep=0.35cm
    \renewcommand{\arraystretch}{1.3}
    \caption{The table presents the objective evaluation of the reconstructed speech from Test-Clean (LibriTTS) with additive term of the modifications to the vanilla snake activation.}
    \vspace{0.1cm}
\begin{tabular}{lcccccc}
\hline
\multicolumn{1}{l|}{Model (MUFFIN)}                 & STFT  & MEL   & PESQ  & STOI  & UTMOS & ViSQOL \\ \hline
\multicolumn{1}{l|}{Added amplitude \& bias (Ours)}                & 1.555 & 0.692 & 2.996 & 0.954 & 4.017 & 4.516  \\
\multicolumn{1}{l|}{Added amplitude}           & 1.603 & 0.744 & 2.928 & 0.945 & 3.943 & 4.448  \\
\multicolumn{1}{l|}{Vanilla}      & 1.635 & 0.760 & 2.876 & 0.940 & 3.905 & 4.409  \\ \hline
\label{tab:act_fn}
\end{tabular}
\end{table}

\section{Illustrations of the auditory feature across various codebook representations.}
\label{appendix:e}
\begin{figure}[!ht]
    \centering    \includegraphics[width=0.85\linewidth]{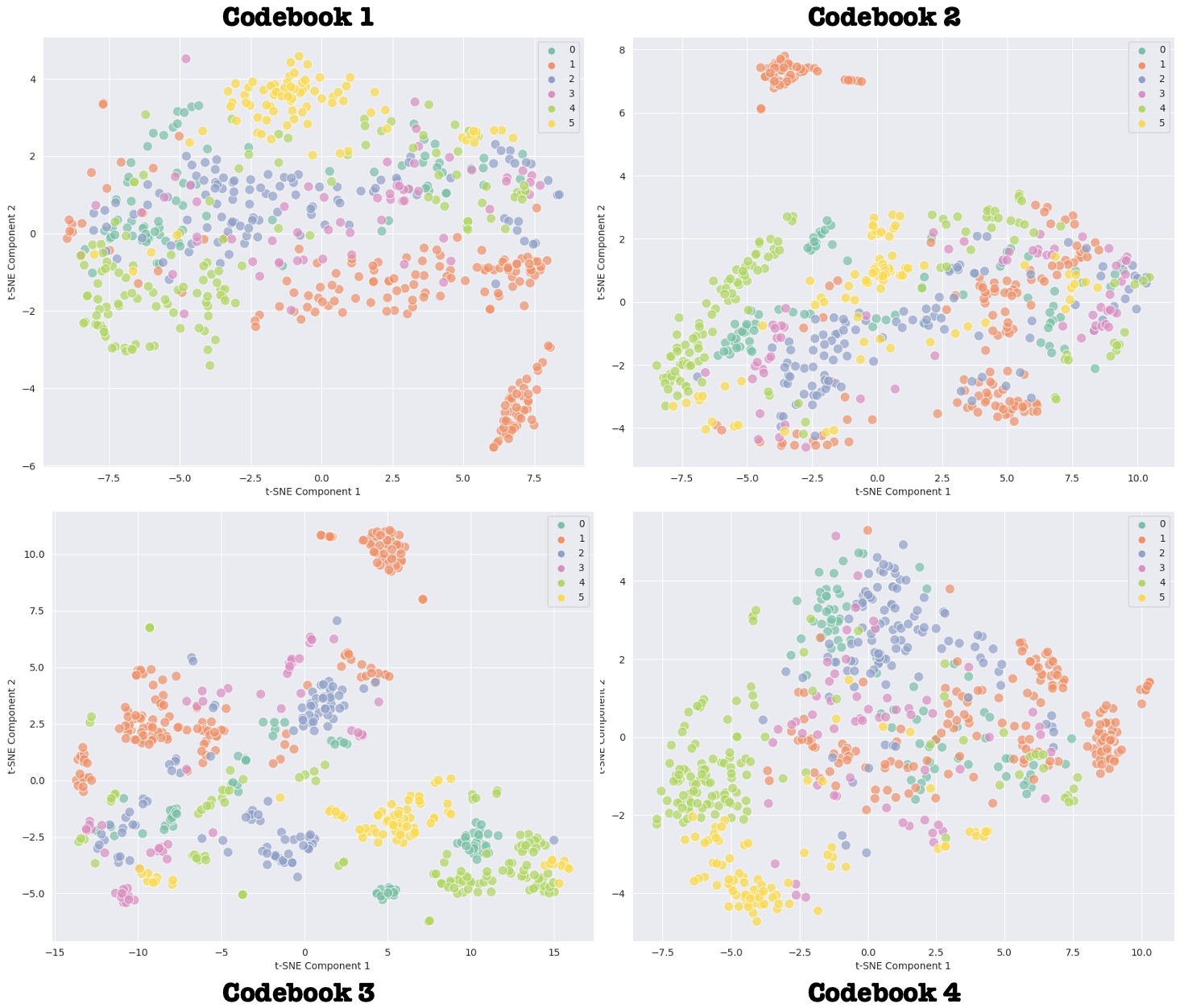}
    \caption{A t-SNE plot showcasing each codebook, with speech randomly sampled from VoxCeleb, effectively represents six distinct speakers of the color code.}
    \label{fig:ch5tsne}
\end{figure}

In this section, we conduct a comprehensive examination of the perceptual characteristics inherent to each distinct quantized frequency band, as revealed by the learned codebooks without target supervision. We note that the cutoff for the frequency band split follows the relative logarithmic scale of the latent sampling rate, according to the psychoacoustic studies. Our analysis utilizes multiple visualization tools to elucidate these characteristics, which include t-SNE plots, randomly sampled spectrograms, and elbow plot. 

These visualizations are derived from the test-clean set of LibriSpeech and a subset of a popular speaker recognition VoxCeleb dataset, allowing us to explore the specific semantic and speaker traits captured within the learned representations. The combined use of these diverse methodologies not only underscores the discriminative power of the representations but also enhances our understanding of their underlying structure and variability. However, it is important to note that the latent average-pooled representations of each sampled utterance were not specifically trained for the speaker recognition (classification) task. Consequently, the optimization objectives did not aim to achieve highly deterministic speaker vectors but rather to encode sequential acoustic and semantic content. This may inherently limits the zero-shot performance of the system on deriving the vector representations for each utterance.

In Figure \ref{fig:ch5tsne}, we justify that among all frequency bands, those in the high-frequency range of 37.5 - 75 Hz distinctly demarcate speaker boundaries with minimal overlap, indicating that \textbf{codebook 3 naturally disentangled speech information to quantize speaker information} without target supervision. This observation is further substantiated by the distances between clusters in the t-SNE plot, where these high-frequency representations are the furthest apart based on the coordinate axis compared to those from the low frequency range of 0 - 18.75 Hz (codebook 1), mid frequency range of 18.75 - 37.5 Hz (codebook 2), and the residuals (codebook 4).

Furthermore, we stress that the reconstructed audio from the demos presented in Section (F) offers compelling evidence that Codebook 3 effectively quantizes speaker attributes from the phonetic content. This is demonstrated by the permutation of the reconstructed audio from different codebooks excluding Codebook 3, is bound to a uniformly flat pitch. Such audio remains invariant to both gender and distinct speaker characteristics, underscoring the unique role of Codebook 3 in capturing these nuances. This resulting outcome can also be used potentially to obtain data samples for normalizing speaker speech, serving the pre-training of applications such as ContentVec \cite{qian2022contentvec}.

\begin{figure}[!ht]
\centering\includegraphics[width=0.83\linewidth]{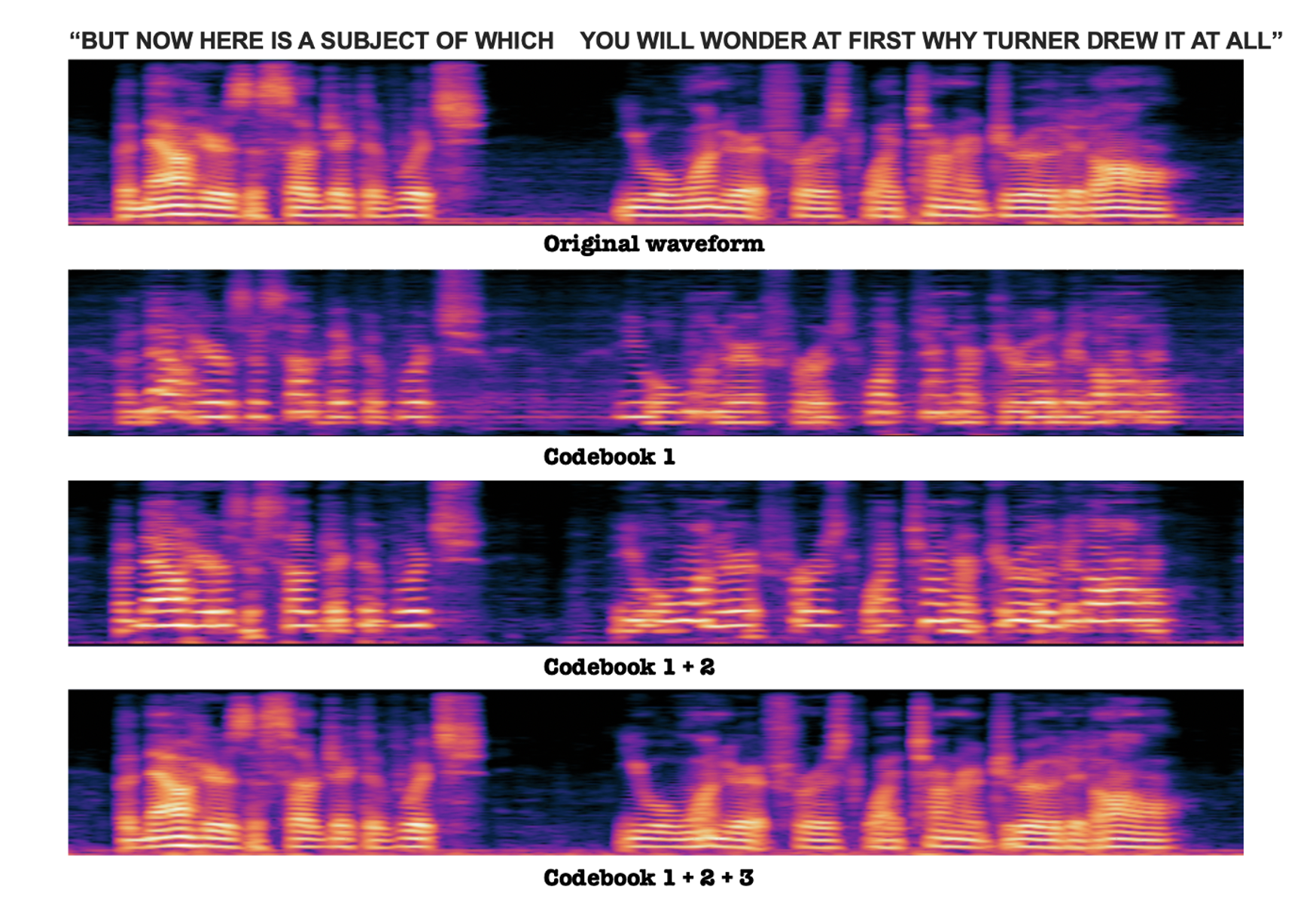}
    \caption{An illustration depicts a randomly sampled speech utterance alongside its reconstruction using incremental codebooks.}
    \label{fig:formant1}
\end{figure}

Likewise, in Figure \ref{fig:formant1}, we present the spectrogram of a randomly sampled speech utterance from the LibriSpeech dataset, alongside its decomposition into different codebooks. This reconstruction incrementally utilizes quantized codebooks. The spectrogram clearly illustrates the previously mentioned flat pitch contours, showing no significant variation from the original waveform when Codebook 3 is omitted from the reconstruction. This effect is evident when only Codebooks 1 and 2 are utilized.

From the spectrogram, we also observe that the frequency of formants is emphasized when combining Codebooks 1 and 2, which reveals clearer phonetic articulation patterns distinguishing vowels and consonants. This observation allows us to appreciate that \textbf{Codebook 2 is responsible for quantizing the articulation and respiratory paralanguage (formant)} of the speech.

\begin{figure}[!ht]
\centering\includegraphics[width=0.65\linewidth]{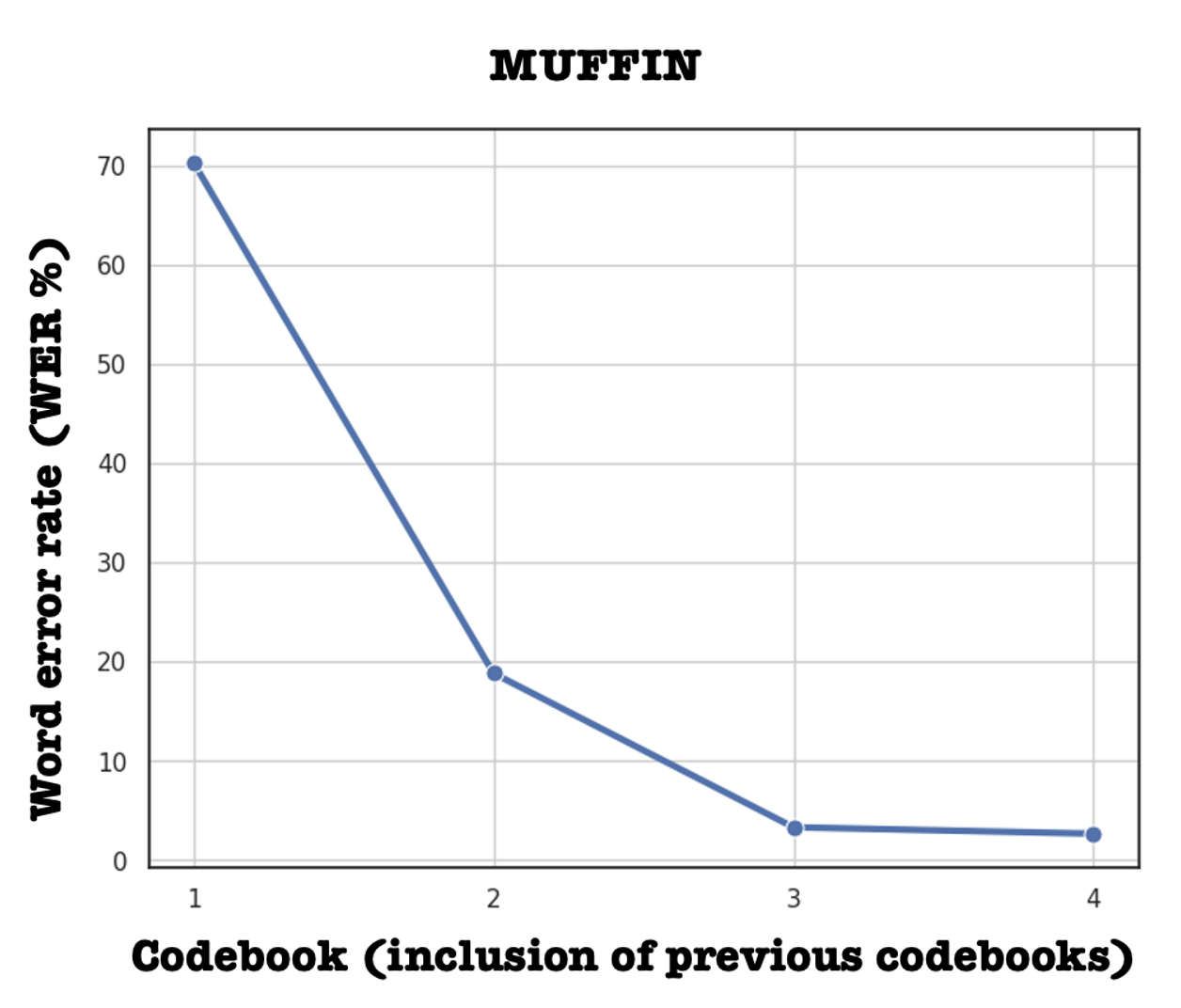}
    \caption{The elbow plot of the word error rate from whisper-large model, utilizing the same setup of incremental codebooks.}
    \label{fig:formant2}
\end{figure}

Next, we plot an elbow curve of the Word Error Rate (WER) as codebooks are added incrementally in Figure \ref{fig:formant2}, illustrating the contribution of each codebook to preserving speech content. A noticeable decrease in the error rate indicates enhanced clarity and intelligibility of speech, attributed to improved articulation. We posit that Codebook 3 does not focus on contextual speech content; using Codebook 3 alone results in high recognition errors, suggesting its limited contribution to core speech intelligibility, while the quantized information in codebook 4 is simply error residual. This analysis, along with auditory demos in Section F, further supports the assertion that Codebooks 1 and 2 are primarily responsible for enhancing speech intelligibility and articulation details.

Consequently, our neural psychoacoustic coding with MUFFIN offers a novel perspective that facilitates the natural disentanglement of speech attributes, guided by label-free psychoacoustic studies. This uniquely positions our approach as an innovative alternative to FACodec \cite{ju2024naturalspeech}, achieving similar goals of obtaining factorized information through a considerably simpler optimization process that could spur further investigation to the advancement of low-resource factorized speech representation learning.

While psychoacoustic studies have primarily focused on speech, applying similar analysis to non-stationary or transient sounds, such as those in music, is both important and intriguing. To explore this, we extended our decomposition approach to a variety of musical genres, including singing, classical, jazz, and symphonic music. Consistent with the psychoacoustic framework used in speech analysis, we observed that:
\begin{itemize}
    \item Codebook 1: primarily captures vocal content and coarse rhythmic beats.
    \item Codebook 2: emphasizes vocal clarity and mid-frequency information.
    \item Codebook 3: encodes pitch details reflective of the singer’s unique characteristics.
\end{itemize}

The above characteristics are demonstrated with samples of music audio presented in our demo page. Interestingly, instrumental content does not clearly separate across Codebooks 2 and 3, suggesting that our psychoacoustic-guided representation is particularly effective in disentangling vocal attributes (speech and singing), but less so for purely instrumental channels. This finding reinforces the theoretical value of psychoacoustic principles for modeling vocal properties, an area that remains underexplored in neural codecs. While applying this framework to instrumental music remains challenging, we believe this opens new research directions. Further investigations, beyond the scope of the current study, may be investigated in our future work.

\section{Discussion on zero-shot text-to-speech synthesis.}

\subsection{Training details}
We evaluate the performance of MUFFIN within the VALL-E framework \cite{wang2023neural}. In contrast to the original study, we train the model with fewer than 600 hours of data. Additionally, during inference, we select a random speech segment from the same speaker to use as the prompt, rather than using the first 1-3 seconds of the speech to be synthesized, as the original study did. This latter approach typically provides more consistent speaker information and constitutes an easier task. These modifications likely account for our lower performance compared to the results reported in the original paper.
%(** argue for the lack of data in training resulting in relatively weaker performance.)

\subsection{Discussion}
During our experiments, we observed that the open-sourced VALL-E configuration did not integrate effectively with MUFFIN operating at a 12.5 Hz sampling rate. This discrepancy highlighted the sensitivity of prompt sequence length in learning sequential decoding information. Longer prompts tended to simplify the task, leading the model to undergeneralize, while shorter prompts provided insufficient information, causing the model to collapse prematurely. Balancing the length of prompts is crucial, particularly as each frame now encapsulates more complex information due to high compression. Consequently, noise in the output sequence reduced speech clarity. Despite these issues, the naturalness of the synthesized speech was reasonably good, demonstrating the benefits of using tokenized audio units. These units separate speech intelligibility from speaker information into distinct, independent codebooks, which are unaffected by adverse conditional computations. This observation emphasizes the importance of the speech codec’s sampling rate in text-to-speech (TTS) systems, directly influencing the quality and intelligibility of synthesized speech. Future work will further investigate this trade-off through systematic experimentation to find optimal configurations that balance resource efficiency with high-quality speech synthesis. 

Furthermore, it would also be highly beneficial to consider weakly supervised training for each psychoacoustic codebook by providing targeted labels in small volumes. Now that we have a clearer understanding of the purpose of each codebook, it becomes more intuitive to apply appropriate supervision with minimal effort, optimizing the quality of the embedding networks while reducing the cost of collecting extensive labeled data for the disentanglement of speech information on factorizing attribute codebook. We anticipate that this will further enhance performance on TTS \cite{zhou2024emotional} or codec-based speech separation \cite{yip2024towards} downstream tasks.

\section{Details of the hyperparameters and specifications of NACs at 24 kHz Sampling Audio}

Note that the MACs (associated with real-time latency) is computed based on a 1-second audio waveform sampled at 24 kHz, using the tool available at https://github.com/sovrasov/flops-counter.pytorch/tree/master. The table below demonstrates that MUFFIN achieves significantly lower MACs, particularly as compression increases with higher downsampling rates, compared to existing codecs. This indicates that MUFFIN offers a lower latency rate than other codecs, including the baseline HiFi-Codec, thereby supporting improved real-time applications.
\begin{table}[!ht]
\centering \footnotesize
\tabcolsep=0.12cm
\renewcommand{\arraystretch}{1.9}
\begin{tabular}{l|ccc|cc|c|c|c|c|c}
\hline
\multirow{2}{*}{Model} &
  \multicolumn{3}{c|}{Num. of Params (M)} &
  \multicolumn{2}{c|}{MACs (G)} &
  \multirow{2}{*}{\begin{tabular}[c]{@{}c@{}}Encoding\\ Rate\end{tabular}} &
  \multirow{2}{*}{\begin{tabular}[c]{@{}c@{}}Downsampling \\ Rate\end{tabular}} &
  \multirow{2}{*}{\begin{tabular}[c]{@{}c@{}}Frame Rate\\ (Hz)\end{tabular}} &
  \multirow{2}{*}{\begin{tabular}[c]{@{}c@{}}Bandwidth\\ (kB/s)\end{tabular}} &
  \multirow{2}{*}{Token/s} \\
           & Encoder & Decoder & Total & Encoder & Decoder &               &      &      &      &     \\ \hline
MUFFIN     & 34.2    & 11.9    & 46.1  & 14.7    & 16.9    & (2, 4, 5, 8)  & 320  & 75   & 2.7  & 300 \\
MUFFIN $\triangledown$    & 34.3    & 11.9    & 46.2  & 5.85    & 8.9     & (4, 5, 6, 8)  & 960  & 25   & 1.35 & 150 \\
MUFFIN $\blacktriangle$     & 36.5    & 14.1    & 50.6  & 6.88    & 11.1    & (3, 5, 8, 16) & 1920 & 12.5 & 0.9  & 100 \\
Encodec    & 7.43    & 7.43    & 14.9  & 1.51    & 4.10    & (2, 4, 5, 8)  & 320  & 75   & 3.0  & 300 \\
DAC        & 21.5    & 52.3    & 73.8  & 18.4    & 64.9    & (2, 4, 5, 8)  & 320  & 75   & 3.0  & 300 \\
Hifi-Codec & 47.2    & 14.3    & 61.5  & 20.6    & 23.8    & (2, 4, 5, 8)  & 320  & 75   & 3.0  & 300 \\ \hline
\end{tabular}
\end{table}
%%%%%%%%%%%%%%%%%%%%%%%%%%%%%%%%%%%%%%%%%%%%%%%%%%%%%%%%%%%%%%%%%%%%%%%%%%%%%%%
%%%%%%%%%%%%%%%%%%%%%%%%%%%%%%%%%%%%%%%%%%%%%%%%%%%%%%%%%%%%%%%%%%%%%%%%%%%%%%%

\end{document}